\documentclass[12pt, draftclsnofoot, onecolumn]{IEEEtran}
\usepackage{amssymb}
\usepackage{amsfonts}
\usepackage{amsmath}
\usepackage{psfrag}
\usepackage{graphicx}
\usepackage{tipa}
\usepackage{subfigure}
\usepackage{bigints}
\usepackage{cite}
\usepackage{xcolor}
\usepackage{array}
\usepackage{multirow}
\usepackage{amsthm}
\usepackage{bm}

\theoremstyle{definition}

\theoremstyle{plain}

\newtheorem{theorem}{Theorem}
\newtheorem{corollary}[]{Corollary}

\newtheorem{proposition}[]{Proposition}

\begin{document}

\title{Efficient UAV Coverage in Large Convex Quadrilateral Areas with Elliptical Footprints}

\author{Alexander~Vavoulas,~Konstantinos~K.~Delibasis,~Harilaos~G.~Sandalidis,~George~Nousias,~and~Nicholas~Vaiopoulos

\IEEEcompsocitemizethanks{\IEEEcompsocthanksitem The authors are with the Department of Computer Science and Biomedical Informatics, University of Thessaly, Papasiopoulou 2-4, 35 131, Lamia, Greece. e-mails\{vavoulas,kdelimpasis,sandalidis,gnousias,nvaio\}@dib.uth.gr.}
}

\maketitle

\begin{abstract}
Unmanned Aerial Vehicles (UAVs) have gained significant attention for improving wireless communication, especially in emergencies or as a complement to existing cellular infrastructure. This letter addresses the problem of efficiently covering a large convex quadrilateral using multiple UAVs, where each UAV generates elliptical coverage footprints based on its altitude and antenna tilt. The challenge is approached using circle-packing techniques within a unit square to arrange UAVs in an optimal configuration. Subsequently, a homography transformation is applied to map the unit square onto the quadrilateral area, ensuring that the UAVs' elliptical footprints cover the entire region. Numerical simulations demonstrate the effectiveness of the proposed method, providing insight into coverage density and optimal altitude configurations for different placement scenarios. The results highlight the scalability and potential for improving UAV-based communication systems, focusing on maximizing coverage efficiency in large areas with
irregular shapes.
\end{abstract}

\begin{IEEEkeywords}
Unmanned aerial vehicles (UAVs), coverage optimization, convex quadrilaterals, ellipse packing, homography transformation, wireless communications
\end{IEEEkeywords}

\section{Introduction}
\IEEEPARstart{U}{nmanned} aerial vehicles (UAVs) can serve as airborne mobile terminals to ensure wireless connectivity during emergencies when cellular networks are unavailable or support terrestrial base stations to improve network performance \cite{J:Vaiopoulos}. Optimizing network coverage and throughput with a fixed number of UAVs presents a practical and intriguing challenge \cite{J:Mozzafari}. For example, in \cite{J:Al-Hourani}, a mathematical model was developed to determine the optimal altitude for a single UAV to maximize its ground service area, while in \cite{J:Nafees}, multi-UAV placement strategies were proposed, utilizing circle-packing techniques.

UAVs can be deployed at varying altitudes, offering increased flexibility in adapting to user distributions and terrain variations. Typically, UAVs are equipped with directional antennas with uniform half-power beamwidths (HPBWs) in both the azimuth and elevation planes, ensuring consistent gains \cite{J:He}. When the antennas are vertically oriented, the resulting footprints are circular. However, circular regions may not accurately reflect actual service areas in practical scenarios. For example, when antennas are tilted at angles relative to the perpendicular surface of the ground, the resulting regions take on an elliptical shape \cite{J:Vavoulas}. The ability to adjust antenna orientations and UAV altitudes allows advanced solutions to optimize the balance between efficiency and resource utilization. These parameters enable dynamic network configurations that address the diverse requirements of ground users under varying environmental conditions \cite{J:Azari}.

In a related study, we addressed the problem of determining the optimal hovering altitude for a single UAV to achieve full coverage of a convex quadrilateral region of arbitrary shape \cite{J:Vavoulas2}. In this letter, we extend the problem to scenarios where the quadrilateral region is sufficiently large, necessitating the deployment of several $M$ UAVs to ensure adequate coverage. Specifically, we focus on covering the quadrilateral using multiple elliptical regions generated by UAVs operating at different altitudes and radiating at varying tilted angles. From a mathematical point of view, the problem is formulated as the packing of a set of arbitrary ellipses, defined by their semi-axes, within a convex polygon \cite{J:Pankratov}. To solve this, we initially apply circle-packing techniques of equal-sized circles within a unit square as a foundational approach. Then, a homography transformation is utilized to map the resulting configurations onto the target quadrilateral region \cite{B:Szabo}, \cite{J:Nousias}. 

Several heuristic iterative methods for packing ellipses into rectangles \cite{J:Kallrath} or regular polygons \cite{J:Pankratov,J:Kampas} have been proposed, but none specifically address packing within arbitrary quadrilaterals. These methods often require significant computational resources and may converge suboptimally, potentially violating non-overlapping constraints. The proposed approach overcomes these challenges by eliminating randomness and iterative optimization processes. It extends the framework presented in \cite{J:Vavoulas2} to a multi-UAV scenario by computing a single $3 \times 3$ projective matrix from a $8 \times 8$ linear system. This enables the construction of multiple ellipses that maintain tangency between their footprints across the entire quadrilateral. Although the methodology in \cite{J:Vavoulas2} could be applied by subdividing the quadrilateral and generating one ellipse per sub-region, this requires solving multiple inscribed ellipse problems. As the number of sub-quadrilaterals increases, tangency and regularity of the UAV layout are not preserved, and the optimal subdivision of the quadrilateral becomes ambiguous.

The following introduces the homography technique that maps the enclosed circles within a unit square onto ellipses within a quadrilateral. Subsequently, a versatile path loss model is applied to determine the optimal placement of the UAVs required to cover the quadrilateral with elliptical footprints. To illustrate the methodology, a numerical example compares two scenarios with different UAV configurations.

\section{Homography Transformation}

\begin{figure}[h]
\centering
\includegraphics[keepaspectratio,width=4in]{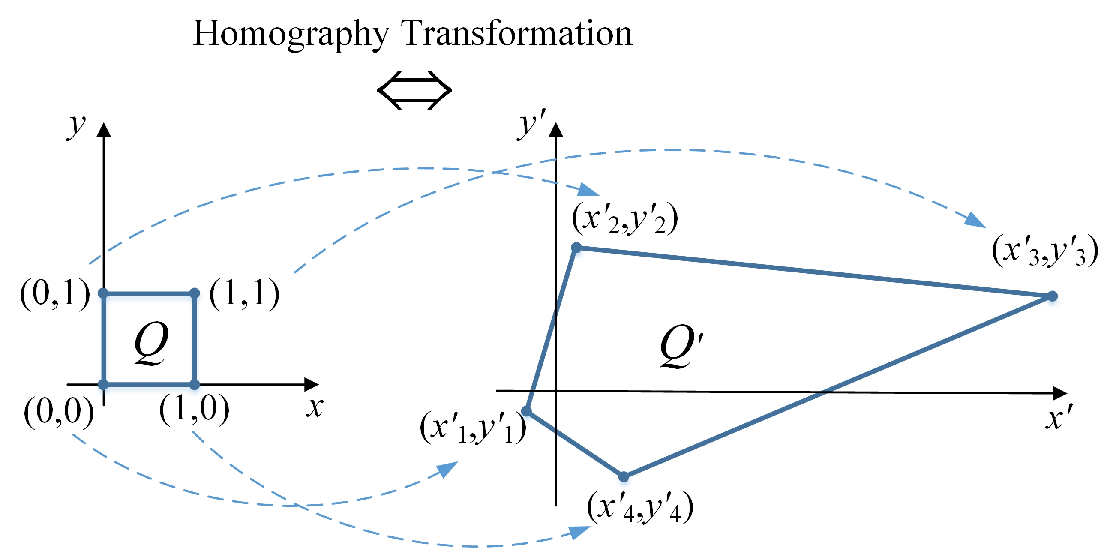}
\caption{ Homography transformation on $Q$.}
\label{Figure1}
\end{figure}

We consider a convex quadrilateral $Q'$ with vertices $P'_i = (x'_i, y'_i)$, for $i = 1, \dots, 4$. The objective is to cover $Q'$ with $M$ elliptical footprints generated by $M$ UAVs. To achieve this, we first define a unit square area $Q$, which contains $M$ circular disks of equal size. In the context of the circle packing problem, the goal is to arrange the $M$ circles within the square to maximize the packing density and ensure that the circles do not overlap. The coordinates of the vertices of $Q$ are given by $P_i = (x_i, y_i)$, for $i = 1, \dots, 4$, where $(x_1, y_1) = (0, 0)$, $(x_2, y_2) = (1, 0)$, $(x_3, y_3) = (1, 1)$, and $(x_4, y_4) = (0, 1)$. Through a homography transformation represented by the matrix $\mathbf{H}$ the square $Q$ is transformed into the quadrilateral the quadrilateral $Q'$, as illustrated in Fig. \ref{Figure1} \cite{B:Hartley}

\begin{gather}
 \begin{bmatrix} x'  \\
y'\\
1 \end{bmatrix}
 =\mathbf{H}
 \begin{bmatrix} x  \\
y\\
1 \end{bmatrix}
=
  \begin{bmatrix}
   h_{11} & h_{12} & h_{13} \\
   h_{21} & h_{22} & h_{23} \\
   h_{31} & h_{32} & h_{33}
   \end{bmatrix}
 \begin{bmatrix} x \\
y\\
1 \end{bmatrix},
\label{HomTrans}
\end{gather}
or equivalently,
\begin{equation}
\begin{aligned}
    x'=\frac{h_{11}x+h_{12}y+h_{13}}{h_{31}x+h_{32}y+h_{33}},~~~
    y'=\frac{h_{21}x+h_{22}y+h_{23}}{h_{31}x+h_{32}y+h_{33}}.
\label{HomTransSys}
\end{aligned} 
\end{equation}
The elements of $\mathbf{H}$ are determined by solving the corresponding homogeneous system of linear equations $\mathbf{B}\cdot \bm{\mathcal{H}}=\mathbf{0}$, where
\begin{equation}
 \bm{\mathcal{H}}^\mathsf{T}=[h_{11} \hspace{4pt} h_{12} \hspace{4pt} h_{13} \hspace{4pt} h_{21} \hspace{4pt}h_{22} \hspace{4pt}h_{23} \hspace{4pt}h_{31} \hspace{4pt}h_{32}\hspace{4pt} h_{33}]  
\end{equation}
and
\begin{gather}
\footnotesize
\hspace{-10pt}\mathbf{B}\hspace{-3pt}=\hspace{-3pt}
\begin{bmatrix}
   -x_1 & -y_1 & -1 & 0& 0 & 0 & x'_1x_1 & x'_1y_1 &x'_1 \\
   0 & 0 & 0 & -x_1 & -y_1 & -1 & y'_1x_1 & y'_1y_1 &y'_1  \\
   -x_2 & -y_2 & -1 & 0& 0 & 0 & x'_2x_1 & x'_2y_2 &x'_2 \\
   0 & 0 & 0 & -x_2 & -y_2 & -1 & y'_2x_2 & y'_2y_2 &y'_1  \\
   -x_3 & -y_3 & -1 & 0& 0 & 0 & x'_3x_3 & x'_3y_3 &x'_3 \\
   0 & 0 & 0 & -x_3 & -y_3 & -1 & y'_3x_3 & y'_3y_3 &y'_1  \\
    -x_4 & -y_4 & -1 & 0& 0 & 0 & x'_4x_4 & x'_4y_4 &x'_4 \\
   0 & 0 & 0 & -x_4 & -y_4 & -1 & y'_4x_4 & y'_4y_4 &y'_1 
\end{bmatrix}. \hspace{-10pt}
\label{MatP}
\end{gather}

The matrix, $\mathbf{H}$, has eight degrees of freedom (DoF) since any nonzero solution to the homogeneous system can be scaled by a constant and still satisfy the system\footnote{Therefore, eight equations are sufficient to determine the nine unknown elements of $\mathbf{H}$.}. Moreover, for a non-degenerate point configuration, i.e., co-linear points, $\mathbf{H}$ is invertible since $\det(\mathbf{H}) \neq 0$. The final solution is determined by the eigenvector of $ \mathbf{B}^\mathsf{T} \mathbf{B} $ associated with the smallest nonzero eigenvalue. This solution satisfies the condition that $\sum_{ij}h_{ij}^2=1$.

The following presents a series of theorems that establish precise mathematical relationships between the mapping of the unit square with the enclosed circles onto the quadrilateral $Q'$ and the ellipses within it. 

\begin{theorem}
The family of horizontal lines $y = y_0$ and vertical lines $x = x_0$, with $x_0$ and $y_0$ as arbitrary constants, intersect under the homography transformation $\mathbf{H}$ at the points $\left( \frac{h_{11}}{h_{31}}, \frac{h_{21}}{h_{31}} \right)$ and $\left( \frac{h_{12}}{h_{32}}, \frac{h_{22}}{h_{32}} \right)$, respectively.
\end{theorem}
\begin{proof}
Consider two horizontal lines in the $\{x,y\}$ plane, $y=y_{i}$ for $i=1,2$. Applying \eqref{HomTransSys}, we obtain
\begin{equation}
x'=\frac{h_{11} x+h_{12} y_{i}+h_{13}}{h_{31} x+h_{32} y_{i}+h_{33}},~~~y'=\frac{h_{21} x+h_{22} y_{i}+h_{23}}{h_{31} x+h_{32} y_{i}+h_{33}}.
\label{xHom}
\end{equation}
By solving the system, the corresponding equations of the lines in the $\{x', y' \}$-plane are deduced

\begin{eqnarray}
y' &=& \frac{h_{31}(h_{23} + h_{22}y_i) - h_{21}(h_{33} + h_{32}y_i)}{h_{31}(h_{13} + h_{12}y_i) - h_{11}(h_{33} + h_{32}y_i)} x' \nonumber \\
&& +\frac{h_{11}(h_{23} + h_{22}y_i) - h_{21}(h_{13} + h_{12}y_i)}{h_{11}(h_{33} + h_{32}y_i) - h_{31}(h_{13} + h_{12}y_i)}.
\label{yHom2}
\end{eqnarray}
Eq. \eqref{yHom2} represents a linear system $2 \times 2$.  The solution provides the intersection point $\{x_{h}',y_{h}'\} =\left\{ \frac{h_{11}}{h_{31}}, \frac{h_{21}}{h_{31}} \right\},$ of the lines with the $\{x', y' \}$-plane. Applying the same procedure to the parallel lines $x = x_{i}$ for $i = 1, 2$, we obtain the intersection point $\{x_{v}',y_{v}'\}=\left\{\frac{h_{12}}{h_{32}}, \frac{h_{22}}{h_{32}}\right\}.$ 
\end{proof}
Note that the above expressions depend solely on $\mathbf{H}$, regardless of $x_0$, or $y_{0}$.

\begin{proposition}
    A homography transformation that maps one convex quadrilateral to another is one-to-one.
\end{proposition}
\begin{proof}
After some algebraic manipulation, the determinant of the Jacobian matrix $\mathbf{J}$ of the homography transformation is obtained as 
\begin{eqnarray}
    \det(\mathbf{J})&=&\frac{h_{31}(h_{12} h_{23}-h_{13} h_{22}) + h_{32}(h_{13} h_{21}  - h_{11} h_{23})}{(h_{33} + h_{31} x + h_{32} y)^3} \nonumber\\ &&+\frac{h_{33}(h_{11} h_{22}- h_{12} h_{21})}{(h_{33} + h_{31} x + h_{32} y)^3}
\end{eqnarray}
Since the nominator does not depend on $x,y$, the determinant is nonzero everywhere in the $\{x,y\}$-plane; thus, the transform is invertible and, therefore, one-to-one \cite{B:Rudin}.
\end{proof}

\begin{theorem}
Under a homography transformation, a circle is mapped to a conic curve.
\end{theorem}
\begin{proof}
The general form of the equation of a circle is given by \cite[eq. (3.315.b)]{B:Bronshtein}
\begin{equation}
    Ax^{2}+Ay^{2}+Dx+Ey+F=0.
\label{CirEq}
\end{equation}
By solving \eqref{HomTransSys} for $x,y$ we get
\begin{equation}
\begin{aligned}
    x=\frac{\hat{h}_{11}x'+\hat{h}_{12}y'+\hat{h}_{13}}{\hat{h}_{31}x'+\hat{h}_{32}y'+\hat{h}_{33}}, ~~
    y=\frac{\hat{h}_{21}x'+\hat{h}_{22}y'+\hat{h}_{23}}{\hat{h}_{31}x'+\hat{h}_{32}y'+\hat{h}_{33}},
\label{HomTransSys2}
\end{aligned} 
\end{equation}
where 
\begin{align*} 
\hat{h}_{11} & = h_{23}h_{32} - h_{22}h_{33}, & \hat{h}_{12} & = h_{12}h_{33} - h_{13}h_{32}, \\
\hat{h}_{13} & = h_{13}h_{22} - h_{12}h_{23}, & \hat{h}_{21} & = h_{21}h_{33} - h_{23}h_{31}, \\
\hat{h}_{22} & = h_{31}h_{13} - h_{11}h_{33}, & \hat{h}_{23} & = h_{11}h_{23} - h_{13}h_{21}, \\
\hat{h}_{31} & = h_{22}h_{31} - h_{21}h_{32}, & \hat{h}_{32} & = h_{11}h_{32} - h_{12}h_{31}, \\
\hat{h}_{33} & = h_{12}h_{21} - h_{11}h_{22}.
\end{align*}

Applying the transformation \eqref{HomTransSys2} to \eqref{CirEq}, the circle is mapped in the $\{x', y' \}$-plane to a conic section, which is generally represented as
\begin{equation}
    A'x'^{2}+B'x'y'+C'y'^{2}+D'x'+E'y'+F'=0,
\label{CirEq2}
\end{equation}
where
\begin{equation}
\begin{split}
A'=&A(\hat{h}_{11}^2+\hat{h}_{21}^2)+\hat{h}_{31}(D\hat{h}_{11}+E\hat{h}_{21})+F\hat{h}_{31}^2, \\
B'=&\hat{h}_{11} \left(2A \hat{h}_{12}+D\hat{h}_{32}\right)+\hat{h}_{21} \left(2A\hat{h}_{22}+E\hat{h}_{32}\right)\\&+\hat{h}_{31} \left(D\hat{h}_{12}+E\hat{h}_{22}+2F\hat{h}_{32}\right),\\
C'=&A(\hat{h}_{12}^2+\hat{h}_{22}^2)+\hat{h}_{32}(D\hat{h}_{12}+E\hat{h}_{22})+F\hat{h}_{32}^2,\\
D'=&\hat{h}_{11} \left(2 A \hat{h}_{13}+D \hat{h}_{33}\right)+\hat{h}_{21} \left(2 A \hat{h}_{23}+E \hat{h}_{33}\right)\\&+\hat{h}_{31} \left(D \hat{h}_{13}+E \hat{h}_{23}+2 F \hat{h}_{33}\right),\\
E'=&\hat{h}_{12} \left(2 A \hat{h}_{13}+D \hat{h}_{33}\right)+\hat{h}_{22} \left(2 A \hat{h}_{23}+E \hat{h}_{33}\right)\\&+\hat{h}_{32} \left(D \hat{h}_{13}+E \hat{h}_{23}+2 F \hat{h}_{33}\right),\\
F'=&A(\hat{h}_{13}^2+\hat{h}_{23}^2)+\hat{h}_{33}(D\hat{h}_{13}+E\hat{h}_{23})+F\hat{h}_{33}^2.
\label{Coeff} \nonumber
\end{split} 
\end{equation}
The conic section is not a circle, as $A' \neq C'$, due to $\det(\mathbf{H}) \neq 0$, which implies $h_{11} \neq h_{21}$, $h_{12} \neq h_{22}$, and $h_{13} \neq h_{23}$.
\end{proof}
 The conditions necessary to represent an ellipse are as follows: $ 4A'C' - {{B'}^2} > 0$ and $C' {{D'}^2} + A' {{E'}^2} - B'D'E' - 4A'C'F' + {{B'}^2}F'> 0$. 
\begin{corollary}
    A circle transformed by a convex-to-convex homography is mapped to an ellipse.
\end{corollary}
\begin{proof}
As stated in Proposition 1, a convex-to-convex homography is one-to-one and invertible. Specifically, the homography in question maps a square to a given quadrilateral. According to Theorem 2, a circle inscribed within the square will be transformed into a conic section that is not a circle. If the transformation results in a hyperbola or parabola, the circle intersects the quadrilateral, which leads to a contradiction. Therefore, the transformation must yield an ellipse.
\end{proof}

\begin{corollary}
    Any two intersecting, non-intersecting, or tangent circles within the square $Q$ are transformed by homography into conic curves that intersect, do not intersect, or are tangent, respectively.
\end{corollary}
\begin{proof}
According to Proposition 1, the homography transformation is one-to-one. Thus, if two curves are tangent at a point, their transformed counterparts will be tangent at a single point since the transformation cannot map the common point to two separate locations. By similar reasoning, any pair of non-intersecting or intersecting curves will be mapped to non-intersecting or intersecting curves, respectively. As a result, if two touching circles share a common point within $Q$, their corresponding conic sections will share the transformed point after homography $\mathbf{H}$.
\end{proof}

The computational aspects of the proposed method can be explicitly determined. The circle-packing layout is taken from deterministic optimal configurations; hence no iterative convergence is required. The homography matrix is computed through a single linear system of equations $8 \times 8$, whereas the UAV altitude optimization presented in the next Section is solved independently for each ellipse. As a result, the total complexity scales linearly with the number of UAVs, $O(M)$, and convergence is guaranteed for all steps. This demonstrates that the method is computationally efficient and scalable.

Since elliptical packing in convex quadrilaterals is unexplored, we compared our method with prior studies in case of rectangles and hexagons. First we prove the following proposition.
\begin{proposition}
    The covered area fraction of $n^2$ identical circles packed in a square is $\frac{\pi}{4}$, which is equal to the covered area fraction of $n^2$ identical ellipses packed in a rectangle via homography transformation.
 \end{proposition}
\begin{proof}
In packing \(n^2\) identical circles with radius \(\frac{w}{2n}\) into a square of side length \(w\), the total coverage area fraction is given by $\frac{1}{w^2}{n^2 \pi \left(\frac{w^2}{4n^2}\right)}=\frac{\pi}{4}$. When this configuration undergoes a homography transformation to a rectangle with side lengths $u$ and $v$, the circles are mapped to $n^2$ identical ellipses. Each ellipse has an area of $\frac{\pi u v}{4n^2}$, thus maintaining the coverage area fraction of $\frac{\pi}{4}$.
\end{proof}

Our method achieved coverage of $\frac{\pi}{4} = 0.7815$ using $n^2$ ellipses in any rectangle, compared to 0.8117 reported in \cite{J:Kallrath} for nine ellipses after 45 minutes of optimization. For a regular hexagon, \cite{J:Kampas} reported 0.8288 with eight ellipses, while our approach, splitting the hexagon into two congruent trapezoids, each packed with four ellipses, yields 0.7715 coverage. Hence, in contrast to the above placement heuristics that are dedicated to these shapes and require iterative optimization, the proposed method achieves similar coverage in a single deterministic step or rectangle and regular polygon ellipse packing.

\section{UAV Placement}
In general, the optimal UAV placement is formulated as the minimization of the maximum path loss, $PL_{\text{max}}$, at the boundary of the coverage region. A simplified path loss model is employed, incorporating the probability of maintaining a line-of-sight (LoS) connection, $\mathbb{P}(\text{LoS})$, influenced by the environment and UAV altitude as defined in \cite{J:Al-Hourani}. The LoS probability is modeled using a sigmoid function with parameters $\eta$ and $\kappa$. At the same time, the elevation angle at the boundary is determined by the UAV height and the geometry of the coverage area \cite{J:Vavoulas2}. The maximum path loss is calculated as a weighted sum of LoS and non-line-of-sight (NLoS) components
\begin{equation}
PL_{\text{max}} = \mathbb{P}(\text{LoS}) \times PL_{\text{LoS}} + \mathbb{P}(\text{NLoS}) \times PL_{\text{NLoS}},
\label{PLmax}
\end{equation}
where $PL_{\text{LoS}}$ and $PL_{\text{NLoS}}$ denote the path losses under LoS/NLoS conditions, respectively,  derived from the total distance $d$, the operating frequency $f$ and additional losses $\xi_{\text{LoS}}$, $\xi_{\text{NLoS}}$ due to scattering and shadowing, and $\mathbb{P}(\text{NLoS})=1-\mathbb{P}(\text{LoS})$\cite{J:Vavoulas2}. This formulation captures the impact of environmental and geometric factors on the optimization of the UAV placement.

Adopting this model for each of the $M$ UAVs deployed to cover the quadrilateral $Q'$, the optimal altitude, \( H_{\text{OPT},i} \) (where $i = 1, \dots, M$), can be determined to minimize the corresponding maximum path loss, \( PL_{\text{max,i}} \). Leveraging the methodology outlined in \cite{J:Vavoulas2}, a unified analytical expression is derived for the optimal altitude of each UAV as follows:
\vspace{-5pt}
\begin{equation}
\small
\begin{aligned}
\hspace{-1pt}PL_{\text{max},i} \hspace{-2pt}= \hspace{-2pt}& \, \frac{\xi_{\text{LoS}} - \xi_{\text{NLoS}}}{1 \hspace{-2pt}+ \eta \exp \left( -\kappa \left( \tan^{-1} \left( \frac{H_{i}b_{i}}{\mathcal{W}} \right) - \eta \hspace{-2pt}\right) \hspace{-2pt}\right)}\hspace{-2pt} & \\
+& 10 \log \left( H_{i}^2\hspace{-2pt} +\hspace{-2pt} \left( \frac{\mathcal{W}}{b_{i}} \hspace{-1pt}\right)^2 \right)\hspace{-2pt}+\hspace{-2pt}20 \log \left( \frac{4 \pi f} {c} \right)\hspace{-2pt} +\hspace{-2pt} \xi_{\text{NLoS}}.
\end{aligned}
\end{equation}
where $\mathcal{W}=a_{i}b_{i} + \sqrt{(b_{i}^2 + H_{i}^2)(a_{i}^2 - b_{i}^2)}$, $\{a_{i}, b_{i}\}$ are the major and minor semi-axes of the elliptical footprint, $H_{i}$ the altitude of the associated UAV and $c$ is the speed of light. The ground projection of each UAV position lies on the major axis of its elliptical footprint. 
The major and minor semi-axes, $a_{i}$ and $b_{i}$, can be expressed through \eqref{CirEq2} as

\begin{equation}
    \{a_{i},b_{i}\}=\sqrt{\frac{\mu}{2} (A'_i+C'_i \pm \sqrt{(A'_i-C'_i)^2+{B'_{i}}^{2}}}) ,
\label{alpha}
\end{equation}
with the $(+)$ sign denoting $a_i$ and the $(-)$ sign denoting $b_i$,  $\mu=4\delta_1\delta_2^{-2}$, $\delta_1=C'_i{D'_i}^2+A_i{E'_i}^2-B'_iD'_iE'_i-F'_i\delta_2$, $\delta_2 = 4A'_i C'_i-B_{i}^2$ for $i = 1, \ldots, 4$. For a given value of $H_i$, the tilt and semi-axis angles, $\psi_i$ and $\theta_i$, respectively, are expressed in terms of $a_i$ and $b_i$ \cite{J:Vavoulas2}
\vspace{-5pt}
\begin{equation}
    \hspace{-6pt}\{\psi_i,\theta_i\}\hspace{-3pt}=\hspace{-3pt}\left\{\hspace{-2pt}\cos^{-1}\hspace{-3pt}\left(\hspace{-2pt}\frac{\sqrt{b_i^2 H_i^2\hspace{-2pt}+\hspace{-2pt}b_i^4}}{\sqrt{a_i^2 H_i^2\hspace{-2pt}+\hspace{-2pt}b_i^4}}\hspace{-2pt}\right), \sin^{-1}\hspace{-3pt}\left(\hspace{-2pt}\frac{b_i^2}{\sqrt{a_i^2 H_i^2\hspace{-2pt}+\hspace{-2pt}b_i^4}}\hspace{-3pt}\right)\hspace{-3pt}\right\}\hspace{-3pt},\hspace{-3pt}
    \label{psi}
\end{equation}

The optimal altitude, $H_{\text{OPT},i}$, for each of the $M$ UAVs can be found by numerically solving for the root of $\frac{\partial PL_{\text{max},i}}{\partial H_{i}}=0.$

\section{Case Study}
We investigated a multi-UAV network operating at a carrier frequency of 2 GHz in different environmental conditions, characterized by the parameter set $(\xi_{\text{LOS}}, \xi_{\text{NLOS}}, \eta, \kappa)$, for the following scenarios: suburban $(0.1, 21, 4.88, 0.43)$, urban $(1, 20, 9.61, 0.16)$, and dense urban $(1.6, 23, 12.08, 0.11)$ \cite{J:Al-Hourani}.  The LoS and NLoS parameters characterize environmental obstruction, with suburban areas exhibiting higher LoS probability and lower loss, and dense urban regions showing greater attenuation.  A typical quadrilateral $Q'$ is considered, preserving the shape shown in Fig. \ref{Figure1}, with vertices at $P'_1 = (-100, -100)$, $P'_2 = (200, -300)$, $P'_3 = (1500, 250)$, and $P'_4 = (50, 400)$ in the $x'$-$y'$-plane, where all coordinates are expressed in meters. Its area can be calculated using the shoelace formula, yielding $S=586,250$m$^2$ \cite{Weisstein}. Without loss of generality, we assume a unit square $Q$. The homography matrix is obtained by substituting the vertices ${x_i, y_i}$ of $Q$ and ${x'_i, y'_i}$ of $Q'$ (for $i=1, \dots, 4$) into $\mathbf{B}$ in \eqref{MatP} and solving the resulting homogeneous system, resulting in
\begin{gather}
\mathbf{H}=
  \begin{bmatrix}
   0.5796  &  0.2807 &  -0.2312 \\
   -0.2912  &  0.6273 &  -0.2312 \\
   -0.0006 & -0.0013 &  0.0023
   \end{bmatrix}.
\label{HomTrans3}
\end{gather}

\begin{figure}[h]
\centering
\subfigure[$M=4$]{\includegraphics[width=4in]{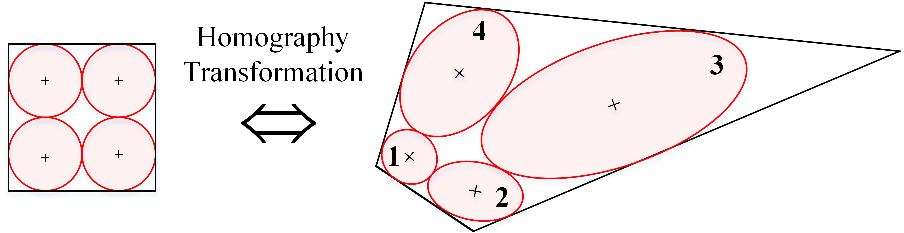}}\\
\subfigure[$M=9$]{\includegraphics[width=4in]{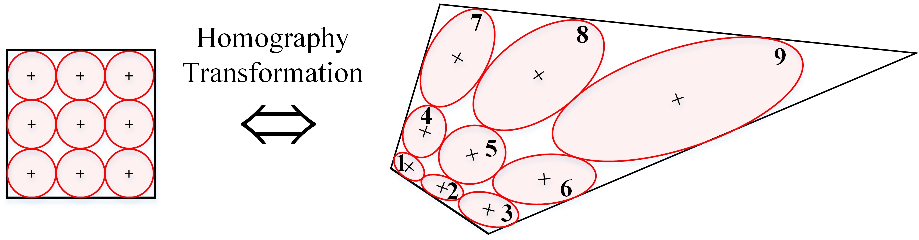}} 
\caption{Circle packing in $Q$ vs. ellipse packing in $Q'$.}
\label{Figure2}
\end{figure}

We focus on two typical UAV configurations consisting of $M=4$ and $M=9$ UAVs. Initially, equal-sized circle packing configurations are applied to the unit square $Q$. These configurations are then transformed into ellipse-packing ones via homography transformation, as depicted in Fig. \ref{Figure2}. The selection of specific $M$ values is determined by their ability to maximize coverage density within the range $M \in [2,15]$ for the circle packing problem, thus optimizing the performance for the quadrilateral area considered \cite{Packomania}. 
Specifically, the total areas of the elliptical footprints are $S_4 = 443,210\text{m}^2$ for the 4-ellipse configuration and $S_9 = 463,426\text{m}^2$ for the 9-ellipse configuration \cite[eq. (3.328a)]{B:Bronshtein}, providing $75.6\%$ and $79.0\%$ coverage of $Q'$, respectively.

\begin{figure}[h]
\centering
\subfigure[suburban]{\includegraphics[width=0.2\textwidth]{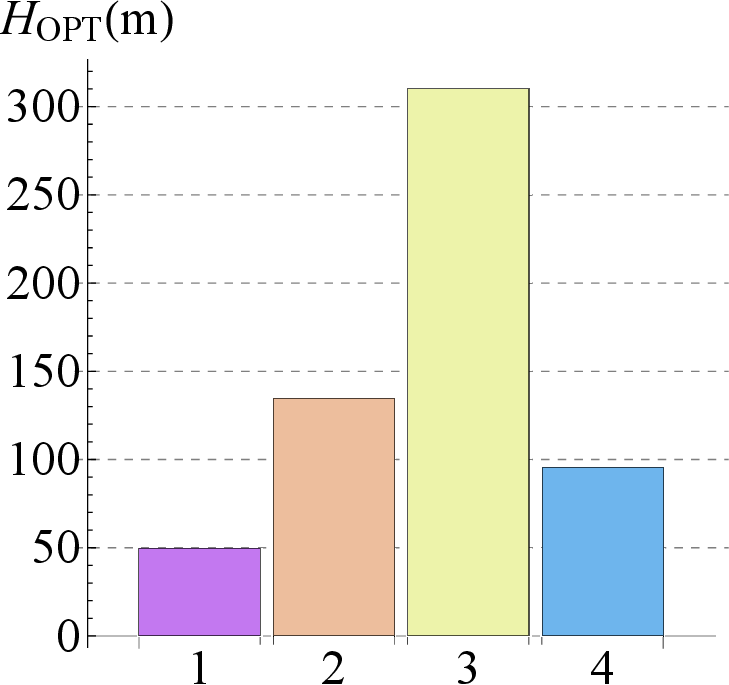}}
\subfigure[urban]{\includegraphics[width=0.2\textwidth]{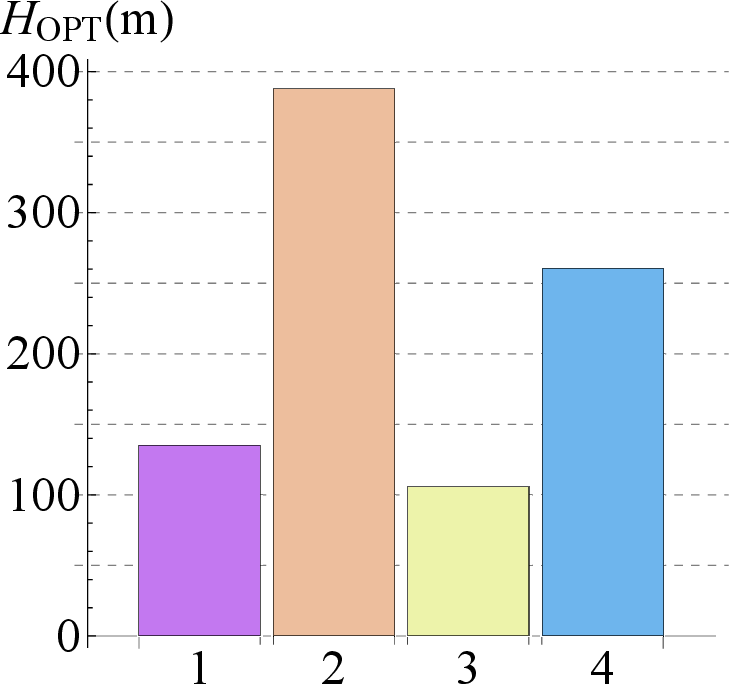}}
\subfigure[dense urban]{\includegraphics[width=0.2\textwidth]{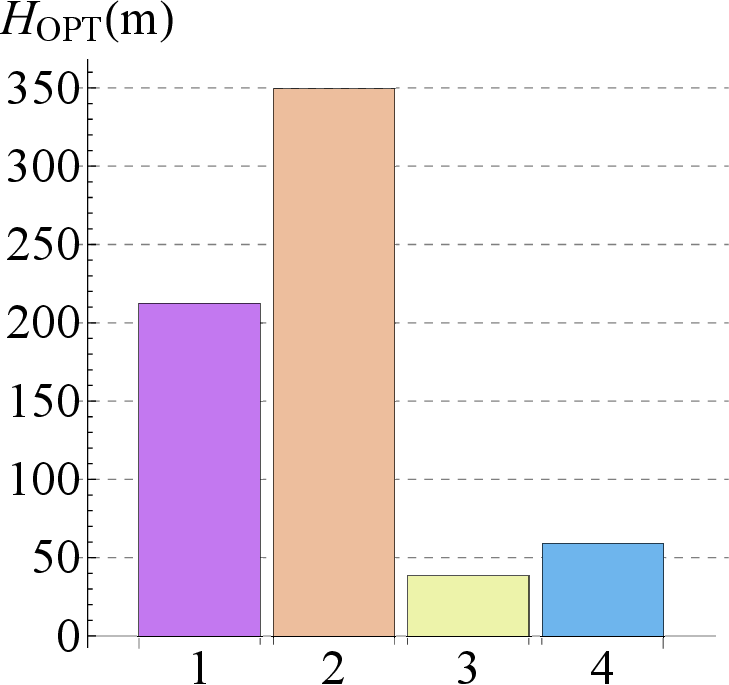}} \\[1ex]
\subfigure[suburban]{\includegraphics[width=0.2\textwidth]{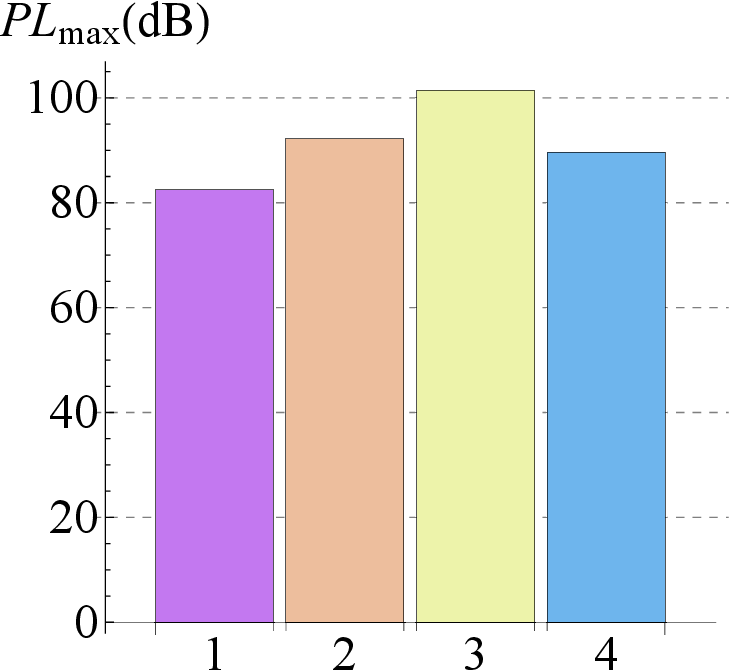}}
\subfigure[urban]{\includegraphics[width=0.2\textwidth]{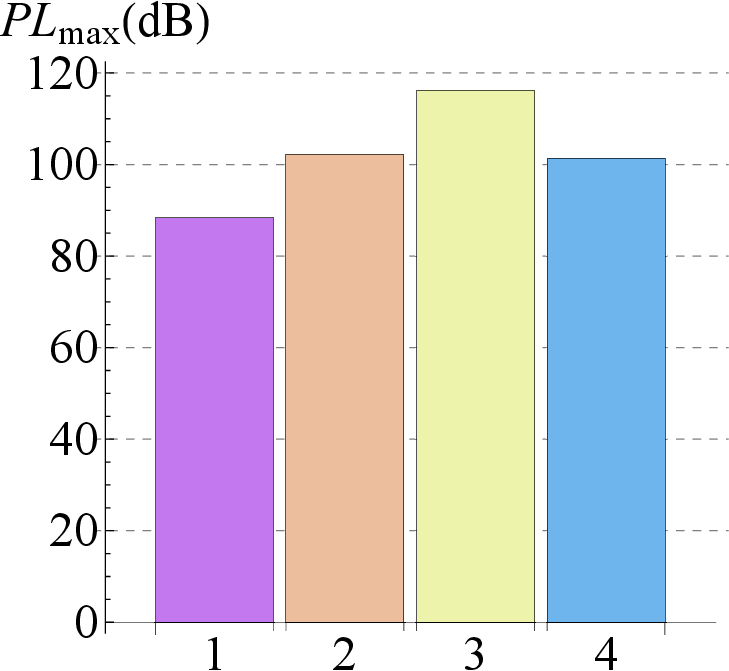}}
\subfigure[dense urban]{\includegraphics[width=0.2\textwidth]{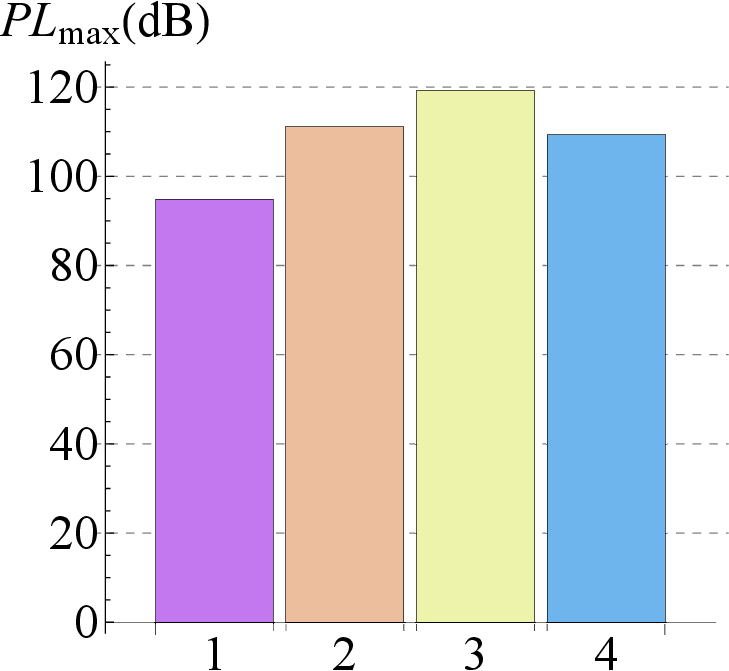}}
\caption{$H_{\mathrm{OPT}}$ and $PL_{\mathrm{max}}$ for four UAVs (indices on the horizontal axes).}
\label{Figure3}
\end{figure}

\begin{figure}[h]
\centering
\subfigure[suburban]{\includegraphics[width=0.2\textwidth]{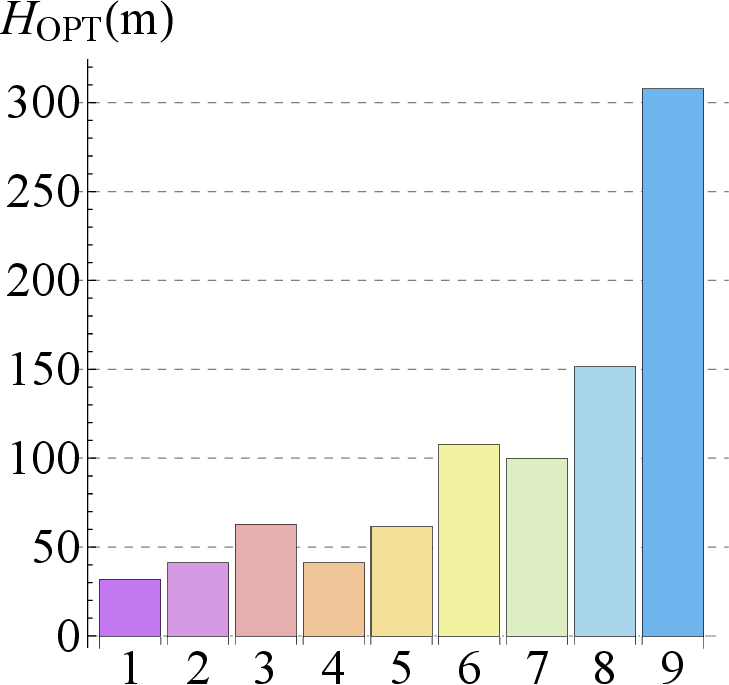}}
\subfigure[urban]{\includegraphics[width=0.2\textwidth]{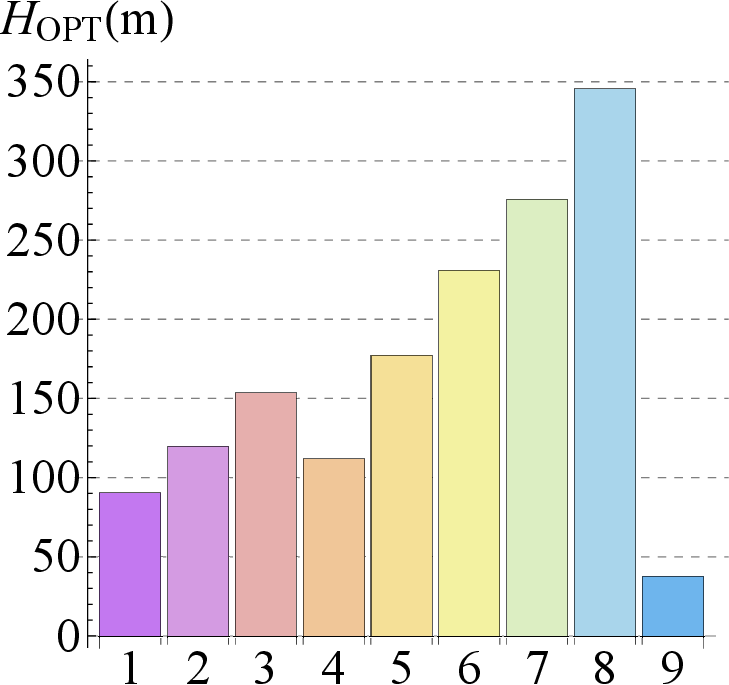}}
\subfigure[dense urban]{\includegraphics[width=0.2\textwidth]{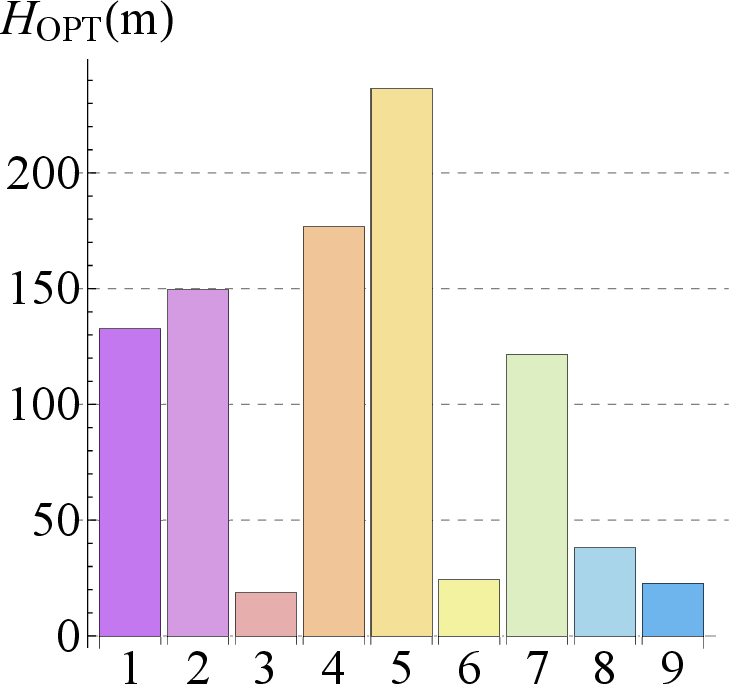}} \\[1ex]
\subfigure[suburban]{\includegraphics[width=0.2\textwidth]{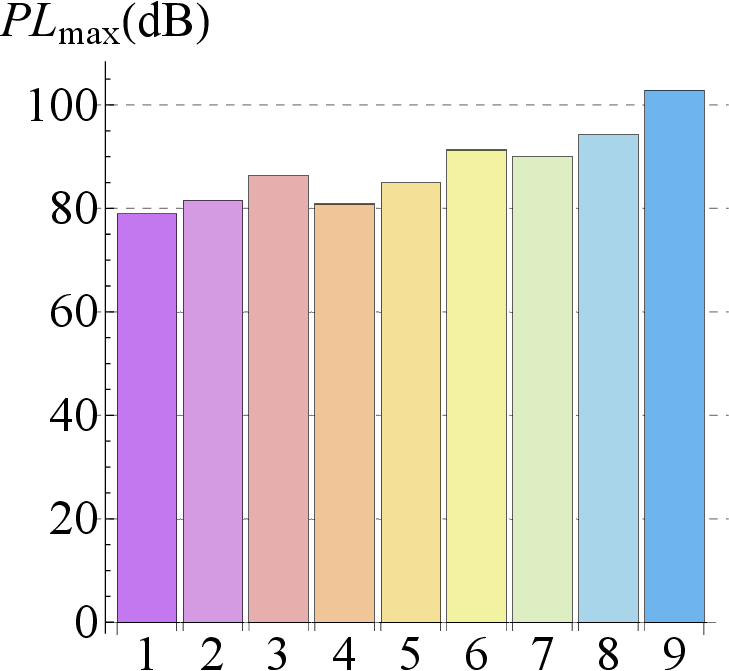}}
\subfigure[urban]{\includegraphics[width=0.2\textwidth]{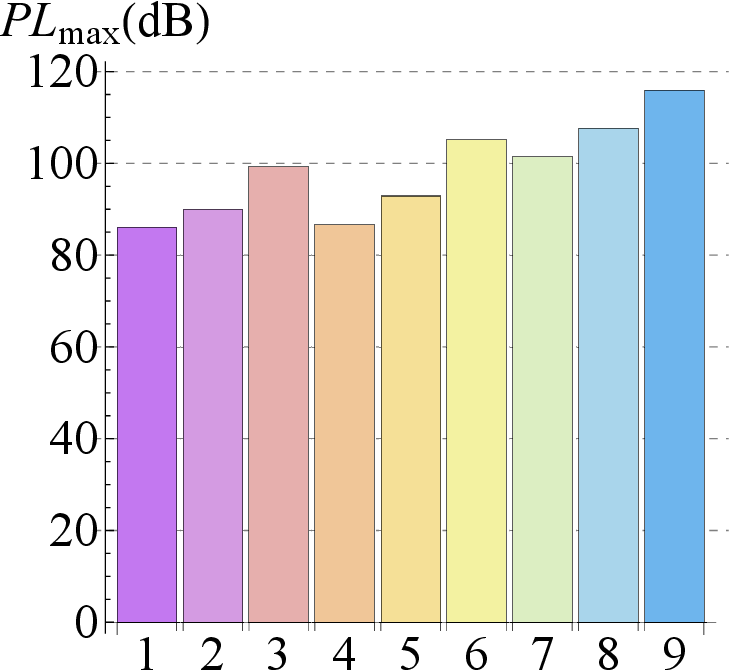}}
\subfigure[dense urban]{\includegraphics[width=0.2\textwidth]{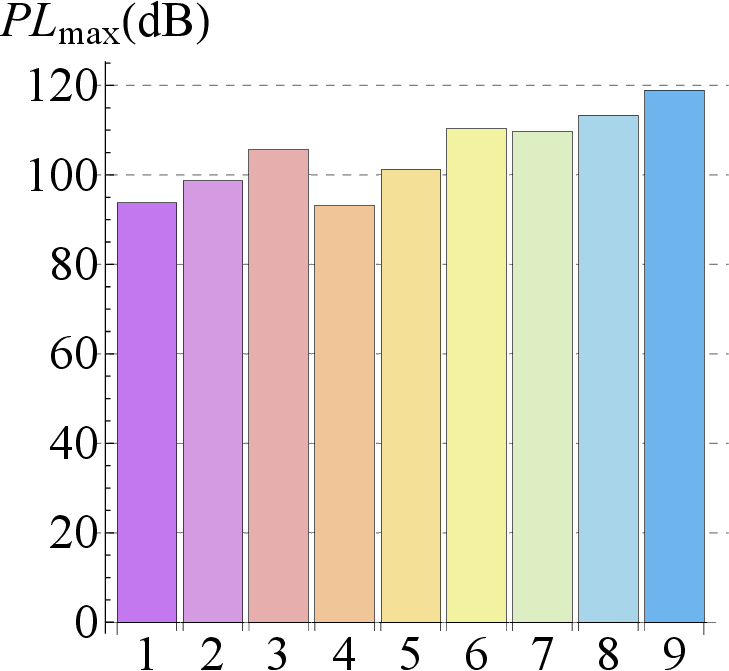}}
\caption{$H_{\mathrm{OPT}}$ and $PL_{\mathrm{max}}$ for nine UAVs (indices on the horizontal axes).}

\label{Figure4}
\end{figure}

Figures \ref{Figure3}, \ref{Figure4} illustrate the optimal altitudes and the corresponding maximum path losses in suburban, urban, and dense urban environments (UAV numbers align with the footprint numbers in Fig. \ref{Figure2}). The results emphasize the flexibility of the variation in altitude between UAVs, which facilitates more robust and adaptive network design strategies. 
The observed variations in $PL_{max}$ offer valuable information to determine the transmit power requirements of each UAV, ensuring the consistent achievement of the minimum quality of service in various operating conditions.

\begin{figure}[h]
\centering
\subfigure{\includegraphics[width=4in]{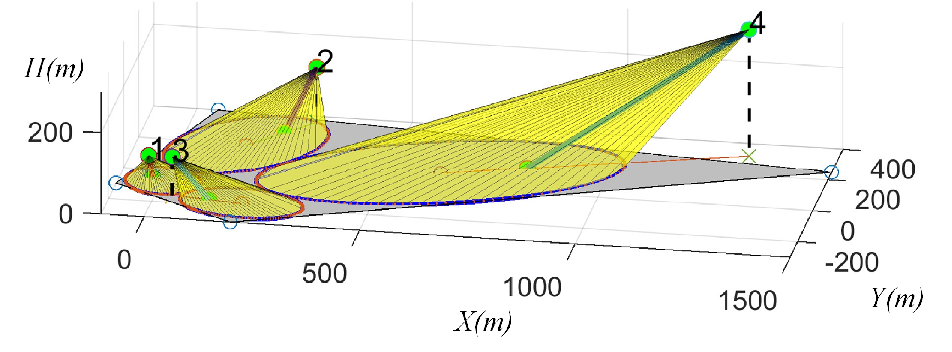}}
\subfigure{\includegraphics[width=4in]{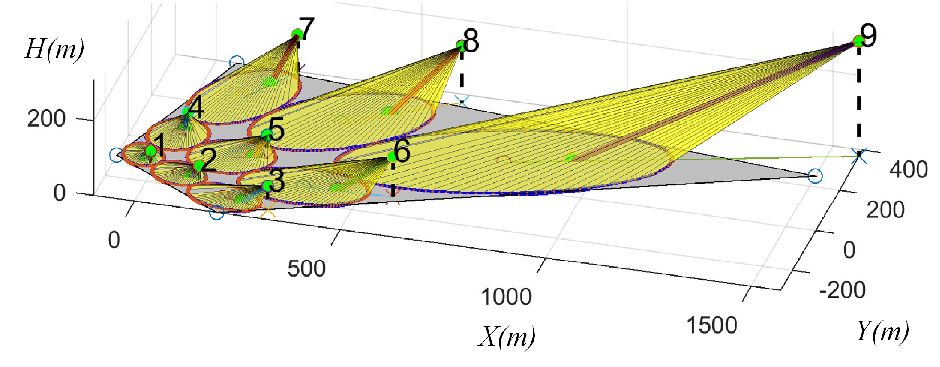}}
\caption{Four and nine UAV placements in the suburban environment.}
\label{Figure5}
\end{figure}

Finally, Fig. \ref{Figure5} demonstrates a detailed three-dimensional placement of the two configurations within a suburban environment. The spatial positions, tilt, and semi-apex angles of the UAVs are depicted, along with the ground footprints.  Additional details regarding the exact semi-axes, optimal altitude, and angle values are included in Table I. Similar graphs can be generated for the other two environments.

\begin{table}[h]
\footnotesize
\centering
\caption{UAV placement metrics for the suburban environment}
\label{Table2}
\renewcommand{\arraystretch}{1.2}
\begin{tabular}{c|c|c|c|c|c|c}
\hline \hline
  & UAV \# & $a_{i}(\text{m})$ & $b_{i}$(\text{m}) & $H_{\text{OPT},i}$(m) & $\theta_i$($^\circ$) & $\psi_i$($^\circ$) \\ 
\hline
  \multirow{4}{*}{\rotatebox{90}{4 UAVs}} & 1 & 93.8 & 83.0 & 49.6 & 56.0 & 15.1 \\ 
  & 2 & 217.0 & 146.8 & 134.7 & 36.4 & 36.3 \\ 
  & 3 & 440.3 & 199.2 & 310.3 & 16.2 & 58.9 \\ 
  & 4 & 149.2 & 91.9 & 95.7 & 30.6 & 42.7 \\ 
\hline
  \multirow{9}{*}{\rotatebox{90}{9 UAVs}} & 1 & 56.5 & 46.0 & 32.0 & 49.5 & 22.2 \\ 
  & 2 & 69.5 & 51.8 & 41.4 & 43.0 & 29.2 \\ 
  & 3 & 95.3 & 54.3 & 62.7 & 26.3 & 47.5 \\ 
  & 4 & 78.7 & 70.2 & 41.3 & 56.6 & 14.4 \\ 
  & 5 & 105.0 & 80.6 & 61.6 & 45.1 & 26.9 \\ 
  & 6 & 161.3 & 86.4 & 108.0 & 23.2 & 50.9 \\ 
  & 7 & 156.5 & 97.7 & 99.9 & 31.4 & 41.8 \\ 
  & 8 & 228.2 & 125.5 & 151.6 & 24.5 & 49.5 \\ 
  & 9 & 413.1 & 155.7 & 308.0 & 10.8 & 65.5 \\ 
  \hline
  \hline
\end{tabular}
\end{table}
 \section{Conclusions and Further Research}
This work addresses the challenge of covering a convex quadrilateral area efficiently using multiple UAVs with elliptical footprints from titled directional antennas. We proposed a  practical UAV placement by applying circle-packing to a unit square and transforming it via homography to fit the quadrilateral. Theoretical analysis and simulations demonstrated that this approach offers competitive coverage, minimizes path loss, and scales effectively with an increasing number of UAVs.

The study can be extended to general convex or non-convex polygons  by decomposing them into smaller convex segments, each mapped using homography transformations. Furthermore, the proposed analysis refers to flat terrain, static UAVs, and an interference-free environment, limiting applicability to dynamic conditions. Future research will address terrain variation, UAV mobility, and interference to improve real-world deployment.

\bibliographystyle{IEEEtran}
\bibliography{IEEEabrv,References}

\begin{thebibliography}{10}
\providecommand{\url}[1]{#1}
\csname url@samestyle\endcsname
\providecommand{\newblock}{\relax}
\providecommand{\bibinfo}[2]{#2}
\providecommand{\BIBentrySTDinterwordspacing}{\spaceskip=0pt\relax}
\providecommand{\BIBentryALTinterwordstretchfactor}{4}
\providecommand{\BIBentryALTinterwordspacing}{\spaceskip=\fontdimen2\font plus
\BIBentryALTinterwordstretchfactor\fontdimen3\font minus \fontdimen4\font\relax}
\providecommand{\BIBforeignlanguage}[2]{{%
\expandafter\ifx\csname l@#1\endcsname\relax
\typeout{** WARNING: IEEEtran.bst: No hyphenation pattern has been}%
\typeout{** loaded for the language `#1'. Using the pattern for}%
\typeout{** the default language instead.}%
\else
\language=\csname l@#1\endcsname
\fi
#2}}
\providecommand{\BIBdecl}{\relax}
\BIBdecl

\bibitem{J:Vaiopoulos}
N.~Vaiopoulos, A.~Vavoulas, and H.~G. Sandalidis, ``An assessment of a unmanned aerial vehicle-based broadcast scenario assuming random terrestrial user locations,'' \emph{IET Optoelectron.}, vol.~15, no.~3, pp. 121--130, 2021.

\bibitem{J:Mozzafari}
M.~Mozaffari, W.~Saad, M.~Bennis, Y.-H. Nam, and M.~Debbah, ``A tutorial on {UAV}s for wireless networks: Applications, challenges, and open problems,'' \emph{{IEEE} Commun. Surveys Tuts.}, vol.~21, no.~3, pp. 2334--2360, 3rd Quart. 2019.

\bibitem{J:Al-Hourani}
A.~Al-Hourani, S.~Kandeepan, and S.~Lardner, ``Optimal {LAP} altitude for maximum coverage,'' \emph{{IEEE} Wireless Commun. Lett.}, vol.~3, no.~6, pp. 569--572, Dec. 2014.

\bibitem{J:Nafees}
M.~Nafees, J.~Thompson, and M.~Safari, ``Multi-tier variable height {UAV} networks: User coverage and throughput optimization,'' \emph{{IEEE} Access}, vol.~9, pp. 119\,684--119\,699, 2021.

\bibitem{J:He}
H.~He, S.~Zhang, Y.~Zeng, and R.~Zhang, ``Joint altitude and beamwidth optimization for {UAV}-enabled multiuser communications,'' \emph{{IEEE} Commun. Lett.}, vol.~22, no.~2, pp. 344--347, Feb. 2018.

\bibitem{J:Vavoulas}
A.~Vavoulas, N.~Vaiopoulos, H.~G. Sandalidis, and K.~K. Delibasis, ``On the terminal location uncertainty in elliptical footprints: Application in air-to-ground links,'' \emph{{IEEE} Trans. Veh. Technol.}, {A}ccepted for publication.

\bibitem{J:Azari}
M.~M. Azari, F.~Rosas, and S.~Pollin, ``Cellular connectivity for {UAVs}: Network modeling, performance analysis, and design guidelines,'' \emph{{IEEE} Trans. Wireless Commun.}, vol.~18, no.~7, pp. 3366--3381, Jul. 2019.

\bibitem{J:Vavoulas2}
\BIBentryALTinterwordspacing
A.~Vavoulas, N.~Vaiopoulos, K.~K. Delibasis, and H.~G. Sandalidis, ``Optimizing coverage in convex quadrilateral regions with a single {UAV},'' 2024. [Online]. Available: \url{https://arxiv.org/abs/2411.18454}
\BIBentrySTDinterwordspacing

\bibitem{J:Pankratov}
A.~Pankratov, T.~Romanova, and I.~Litvinchev, ``Packing ellipses in an optimized convex polygon,'' \emph{J. Glob. Optim.}, vol.~75, no.~3, pp. 495--522, 2019.

\bibitem{B:Szabo}
P.~G. Szabó, M.~C. Markót, T.~Csendes, E.~Specht, L.~G. Casado, and I.~García, \emph{New Approaches to Circle Packing in a Square}.\hskip 1em plus 0.5em minus 0.4em\relax New York, NY, USA: Springer-Verlag, 2007.

\bibitem{J:Nousias}
G.~Nousias, K.~Delibasis, and I.~Maglogiannis, ``Intelligent sampling consensus for homography estimation in football videos using featureless unpaired points,'' \emph{{IEEE} Access}, vol.~13, pp. 187\,843--187\,857, 2025.

\bibitem{J:Kallrath}
J.~Kallrath and S.~Rebennack, ``Cutting ellipses from area-minimizing rectangles,'' \emph{J. Glob. Optim.}, vol.~59, no.~2, pp. 405--437, 2014.

\bibitem{J:Kampas}
F.~J. Kampas, I.~Castillo, and J.~D. Pint{\'e}r, ``Optimized ellipse packings in regular polygons,'' \emph{Optim. Lett.}, vol.~13, no.~7, pp. 1583--1613, 2019.

\bibitem{B:Hartley}
R.~Hartley and A.~Zisserman, \emph{Multiple View Geometry in Computer Vision}, 2nd~ed.\hskip 1em plus 0.5em minus 0.4em\relax New York, NY, USA: Cambridge Univ. Press, 2003.

\bibitem{B:Rudin}
W.~Rudin, \emph{Principles of Mathematical Analysis}, 3rd~ed.\hskip 1em plus 0.5em minus 0.4em\relax New York, NY, USA: McGraw-Hill (Inc.),, 1976.

\bibitem{B:Bronshtein}
I.~N. Bronshtein, K.~A. Semendyayev, G.~Musiol, and H.~M{\"u}hlig, \emph{Handbook of Mathematics}, 6th~ed.\hskip 1em plus 0.5em minus 0.4em\relax New York, NY, USA: Springer-Verlag, 2015.

\bibitem{Weisstein}
\BIBentryALTinterwordspacing
E.~W. Weisstein, ``Shoelace formula,'' From MathWorld--A Wolfram Web Resource, 2024, accessed: 2024-11-19. [Online]. Available: \url{https://mathworld.wolfram.com/ShoelaceFormula.html}
\BIBentrySTDinterwordspacing

\bibitem{Packomania}
\BIBentryALTinterwordspacing
Packomania, ``Packomania: The circle packing problem,'' 2024, accessed: 2024-12-21. [Online]. Available: \url{http://www.packomania.com/}
\BIBentrySTDinterwordspacing

\end{thebibliography}

\end{document}